\documentclass[reprint,amsmath,amssymb,aps,pre,longbibliography,superscriptaddress,showkeys]{revtex4-2}
\usepackage[english]{babel}
\usepackage{graphicx}
\usepackage{amsfonts}
\usepackage{amsmath}
\usepackage{svg}
\usepackage{svg-extract}
\usepackage{ragged2e}
\usepackage{hyperref}
\usepackage{titlesec}
\titlespacing*{\subsection}{0pt}{\baselineskip}{0.5\baselineskip}

\newcommand{\E}{{\mathbb{E}}}

\DeclareMathOperator{\Var}{Var}

\begin{document}

\title{Phase transitions in microbial lineage trees}

\author{Kaan \"Ocal}
\email{kaan.ocal@unimelb.edu.au}
\affiliation{%
School of BioSciences, University of Melbourne, Parkville, Victoria 3010, Australia
}%
\affiliation{%
School of Mathematics and Statistics, University of Melbourne, Parkville, Victoria 3010, Australia
}%
\author{Syrine Ghrabli}
\affiliation{%
School of BioSciences, University of Melbourne, Parkville, Victoria 3010, Australia
}%
\author{Michael P.H. Stumpf}
\affiliation{%
School of BioSciences, University of Melbourne, Parkville, Victoria 3010, Australia
}%
\affiliation{%
School of Mathematics and Statistics, University of Melbourne, Parkville, Victoria 3010, Australia
}%
\affiliation{%
Cell Bauhaus Pty Ltd, Carlton, Victoria 3052, Australia
}%

% Statistical physics can describe the behavior of microbial populations consisting of many heterogeneous individuals. A direct consequence is the existence of phase transitions, where the behavior of a population changes discontinuously upon a small perturbation. While such phase transitions have often been proposed in biology, connecting observed behavior to the underlying physics has remained challenging. We show how phase transitions naturally arise in microbial population dynamics and highlight their connection with genealogies. We rigorously demonstrate the existence of a first-order phase transition in a model of bacterial plasmid engineering and find a strict lower bound on the number of plasmids that can be stably maintained in a population. 

\begin{abstract}
Microbial populations exhibit high cell-to-cell variability, which fundamentally shapes population behavior. A striking consequence is the existence of phase transitions, where small genetic or environmental changes trigger abrupt shifts in population dynamics. While biological phase transitions have often been proposed, connecting observed behavior to the underlying physics has remained challenging. We combine population genetics with statistical physics to show how phase transitions arise naturally in microbial populations. We highlight the existence of a first-order transition in a model of bacterial plasmid engineering and find a strict lower bound on the number of plasmids that can be stably maintained in a population. 
\end{abstract}


\maketitle 

\section{Introduction}

% Old version:
%Ideanotes on improving intro readability: 1.Anchor to Kaan's previous work. 1bis.A faster jump into the topic perhaps?  2."Ideas from statistical physics can explain" is too nuanced/broad. Replace with one example, one that is easy/simple/generic. 3.Forgo airy terms like "naturally" that makes it feel like something mysterious and magic is happening behind. 4.I think the most opaque sentence to me is "This relies on fundamental symmetries in the large-deviation behavior of counting processes and their associated first-passage times"
    
Microbial populations are characterized by extensive cell-to-cell variability caused e.g.~by noise in gene expression and inheritance, and environmental fluctuations \cite{piho_feedback_2024,stine_lineage-dependent_2025,levien_non-genetic_2021}. An important question is how differences in individual cells affect population behavior on a macroscopic scale \cite{levien_non-genetic_2021}. In this paper we study how single-cell variability can cause threshold phenomena, where small perturbations to a population trigger abrupt shifts in behavior. In statistical physics, such threshold phenomena are typically explained by phase transitions; observations consistent with phase transitions have previously been proposed in population biology \cite{anderson_more_1972,sole_phase_1996,munoz_colloquium_2018,ordway_phase_2020,heffern_phase_2021,bhattacharyya_emergence_2024}, but there have been few decisive demonstrations of their existence \cite{bagchi_phase_2011,skanata_evolutionary_2016,mayer_transitions_2017,lin_optimal_2019}. In this paper, we show how first order phase transitions arise naturally in microbial populations and demonstrate their relevance in bacterial bioengineering.

Our work builds on the approach to microbial population dynamics developed in \cite{wakamoto_optimal_2012,nozoe_inferring_2017,garcia-garcia_linking_2019,kobayashi_fluctuation_2015,sughiyama_pathwise_2015}, which represents a population as a collection of interrelated lineages. The large-scale behavior of these lineages can be understood via large deviation theory \cite{dembo_large_2010,levien_large_2020}, the mathematical formalism underlying much of modern statistical physics \cite{touchette_large_2009}. This leads to a natural thermodynamic correspondence where the growth rate of a population plays a role exactly analogous to free energy in classical physics \cite{pigolotti_generalized_2021,ocal_two-clock_2025}. We show that singularities in the growth rate represent phase transitions where certain population quantities change discontinuously, with the connection being given by the Euler-Lotka equation. At first-order transitions, these population quantities can be represented as selection coefficients, directly tying the large deviation perspective to population genetics.

% %Syrine: 
%  % Examples of similar work include directed percolation in something (the paper you want to cite), and one of examples from MS papers sent, but one can extend further principles from dynamic phase transition theory \cite{DPT} to time-dependent systems exhibiting singularities with direct analogs of thermodynamic potentials, free energy and entropy (cite). This use of analogy allows for new results in population genetics and out-of-equilibrium dynamics phase transitions alike.

% Biological cells across the tree of life share three characteristics: they communicate with their environment, including other cells; the consume resources and produce energy and biomass to sustain their physiology and often (not all cell-types) to replicate; and they adjust their physiological and metabolic state in response to internal and external cues. In biotechnology their ability to turn substrate into product can be harnessed to produce small molecules, lipids, proteins, and biopolymers. Some 50\% of products currently being produced using petrochemistry could be generated this way, typically with vastly reduced environmental burdens.

We demonstrate our results on a simple model of precision fermentation, which uses engineered plasmids in bacteria to produce biomolecules at industrial scale. Here a commonly observed difficulty is that synthetic plasmids interfere with their hosts' normal functions and decrease their evolutionary fitness \cite{rugbjerg_synthetic_2018}. This together with the highly stochastic nature of plasmid transmission \cite{hernandez-beltran_segregational_2022} typically results in plasmids ultimately being lost from the population, negating the synthetic capability of engineered cells. We show that the genetic drift arising from imperfect plasmid segregation leads to a discontinuous phase transition between survival and extinction, and imposes a lower limit on the number of plasmids per cell that can be stably maintained in a population: unless strong evolutionary pressure is applied, populations can only maintain high ($>10$--$15$) copies of plasmids per cell or none at all. Our example sheds light on how stochastic copy number control and partitioning in plasmids affect their long-term viability. % (concluding remarks by Michael)

% This gets directly into finite size effects that I'd rather not have to think about
% The problem of escape mutants tends to become visible once a strain moves from the lab to pilot and production scales, i.e. $\gtrsim 50$~l or $\gtrsim 1,000$~l, respectively. Population dynamics and the relative trade-off between selection and random demographic effects will shift considerably between these scales. Here more is really different and the phase transition highligthed here reflects the increasing dominance of selection at large population sizes.

% Paragraph 1:

% %The stochastic micro / deterministic ergodic argument open problem; I'd hedge my bets when using statphys to results w/ no empirical / proof

% % Beyond the traditional thermodynamic systems, biological systems with large stochastic degrees of freedom make great candidates where statistical physics is now used to derive quantitative laws for (evolutionary/pop/gen) outcomes

% %No phase transitions in finite systems (finite N and finite t)
% Selective pressures can be measured!
% Quote: "suggestive but not uniquely diagnostic of a thermodynamic phase transition"
%syrine: elighten me on : Wild type, perturbation theory, culumant/proba, order/states/, concepts
% We show that phase transitions arise naturally as an intrinsic property of biological populations and do not require environmental fluctuations.

% (PT in a three-species ecological model), \cite{heffern_phase_2021}, \cite{scott_phase_2022} (review)% Focus on second-order phase transitions

\section{Thermodynamics of lineages}

\begin{figure}[t]
    \centering
    %% Creator: Inkscape 1.4.4 (dcaf3e7d9e, 2026-05-05), www.inkscape.org
%% Accompanies image file 'lineages.pdf' (pdf, eps, ps)
\begingroup%
  \makeatletter%
  \providecommand\color[2][]{%
    \errmessage{(Inkscape) Color is used for the text in Inkscape, but the package 'color.sty' is not loaded}%
    \renewcommand\color[2][]{}%
  }%
  \providecommand\transparent[1]{%
    \errmessage{(Inkscape) Transparency is used (non-zero) for the text in Inkscape, but the package 'transparent.sty' is not loaded}%
    \renewcommand\transparent[1]{}%
  }%
  \providecommand\rotatebox[2]{#2}%
  \newcommand*\fsize{\dimexpr\f@size pt\relax}%
  \newcommand*\lineheight[1]{\fontsize{\fsize}{#1\fsize}\selectfont}%
  \ifx\svgwidth\undefined%
    \setlength{\unitlength}{229.23890602bp}%
    \ifx\svgscale\undefined%
      \relax%
    \else%
      \setlength{\unitlength}{\unitlength * \real{\svgscale}}%
    \fi%
  \else%
    \setlength{\unitlength}{\svgwidth}%
  \fi%
  \global\let\svgwidth\undefined%
  \global\let\svgscale\undefined%
  \makeatother%
  \begin{picture}(1,0.58958575)%
    \lineheight{1}%
    \setlength\tabcolsep{0pt}%
    \put(0,0){\includegraphics[width=\unitlength,page=1]{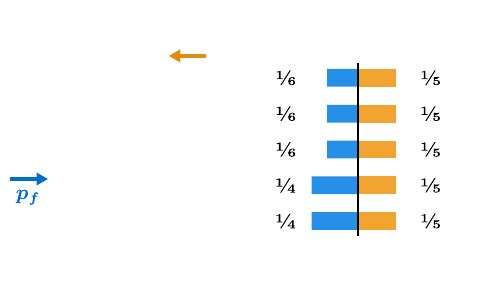}}%
    \put(0.59740031,0.55797644){\color[rgb]{0,0,0}\transparent{0.89999998}\makebox(0,0)[t]{\smash{\begin{tabular}[t]{c}forward\\probability\end{tabular}}}}%
    \put(0.90066722,0.55797644){\color[rgb]{0,0,0}\transparent{0.89999998}\makebox(0,0)[t]{\smash{\begin{tabular}[t]{c}backward\\probability\end{tabular}}}}%
    \put(0.27214674,0.01912983){\color[rgb]{0,0,0}\transparent{0.89999998}\makebox(0,0)[t]{\smash{\begin{tabular}[t]{c}time\end{tabular}}}}%
    \put(0.49975038,0.02856204){\color[rgb]{0,0,0}\transparent{0.89999998}\makebox(0,0)[t]{\smash{\begin{tabular}[t]{c}$t$\end{tabular}}}}%
    \put(0,0){\includegraphics[width=\unitlength,page=2]{lineages.pdf}}%
  \end{picture}%
\endgroup%

    \caption{\justifying  A population consisting of interrelated lineages. The forward distribution $p_f$ follows a lineage in time, starting with the original ancestor and selecting a random offspring at each reproduction event. The forward probability of a lineage depends on its multiplicity in each generation via Eq.~\eqref{eq:pop_fwd_prob}. The backward distribution $p_b^t$ assigns the same probability to the ancestral lineage, or genealogy, of each individual at time $t$.}
    \label{fig:intro}
\end{figure}

We identify each cell in a population with its ancestral lineage, and therefore view a population as a tree made up of interrelated lineages \cite{stadler_phylodynamics_2021} (Fig.~\ref{fig:intro}). For simplicity we assume that all lineages stem from a common ancestor. The $i$-th individual is born at time $t_i$ and is one of $m_i$ offspring, where $t_1 = 0$ and $m_1 = 1$ for the first individual. We define the generation counting process for a lineage as
\begin{align}
    n(t) &= i \qquad (t_i \leq t < t_{i+1}). \label{eq:pop_counting_process}
\end{align}

\noindent Here we assume that each individual dies upon reproduction, which captures the microbial case; we also assume that every individual will eventually reproduce and neglect extinction. These assumptions do not fundamentally alter our results \cite{genthon_cell_2023,ocal_two-clock_2025}.

We can relate lineage and population statistics by means of the \emph{forward distribution} over lineages \cite{nozoe_inferring_2017}. The forward distribution is defined as follows: starting from the universal ancestor, go down the population tree by picking a descendant uniformly at random in each generation. Since there are $m_i$ descendants in the $i$-th generation, each descendant has probability $m_i^{-1}$ of being chosen. The forward probability of a lineage $\ell$ at time $t$ therefore equals
\begin{align}
    p_f(\ell) = m_1^{-1} \cdots m_{n(t)}^{-1}. \label{eq:pop_fwd_prob}
\end{align}

\noindent We recover the population size $N(t)$ at time $t$ as
\begin{align}
    N(t) &= \E_f\left[ \prod_{i=1}^{n(t)} m_i \right].  \label{eq:pop_nozoe}
\end{align}

\noindent To see this, note that the product in \eqref{eq:pop_nozoe} exactly cancels out the forward probability \eqref{eq:pop_fwd_prob}, so every lineage contributes $1$ to the expectation. We therefore define the weight of a lineage as
\begin{align}
    w(t) &= \prod_{i=1}^{n(t)} m_i, \label{eq:pop_abs_fitness}
\end{align}

\noindent which biologically represents the absolute fitness of a lineage, i.e.~its contribution to the population.

% The above argument shows that we can express population statistics at time $t$ using forward lineages. If $x$ is a variable associated to every cell, such as the number of plasmids in a bioengineering context, then we can write
% \begin{align}
%     \E_p[ x(t) ] &= \frac{\E_f\left[ w(t, \ell) x(t, \ell) \right]}{\E_f\left[ w(t, \ell) \right]}
% \end{align

% \noindent Here the weight $w(t, \ell)$ ensures that each lineage that is alive at time $t$ contributes $1$, and each lineage that is extinct by time $0$ contributes $0$.

In the thermodynamic framework of \cite{levien_large_2020}, we treat each lineage of the population as a possible outcome or \emph{state} of our system. The population size $N(t)$ then counts the number of states and represents the partition function, typically denoted by $Z$. With \eqref{eq:pop_nozoe} we get for the population growth rate  
\begin{align}
    N(t) \sim e^{\Lambda t} &\implies \Lambda = \lim_{t \rightarrow \infty} \frac 1 t \log \E_f[ w(t) ]. \label{eq:lambda_def}
\end{align}

\noindent Here time $t$ is a proxy for the system size, and the thermodynamic limit corresponds to letting $t \rightarrow \infty$. In \eqref{eq:lambda_def} the growth rate $\Lambda$ takes the role of the free energy density (up to sign) \cite{skanata_evolutionary_2016}, or the normalized log-partition function $\log Z$. A population favors lineages that maximize the growth rate, much like a thermodynamic system favors states that minimize the free energy. 

Let $x(t)$ be a time-dependent observable in a lineage and define the time integral
\begin{align}
    X(t) &:= \int_0^t x(s) \, ds.
\end{align}

\noindent To analyze the large-scale behavior of $x(t)$ in a population, we define the rescaled cumulant generating function
\begin{align}
    \phi(\lambda) &= \lim_{t \rightarrow \infty} \frac 1 t \log \E_f[ w(t) \, e^{\lambda X(t)}]. \label{eq:pf_def}
\end{align}

\noindent This reduces to the log-partition function in \eqref{eq:lambda_def} for $\lambda = 0$. Differentiating with respect to $\lambda$ (see Apdx.~\ref{apdx:cgf}) yields
\begin{align}
    \phi'(0) &=\lim_{t \rightarrow \infty} \E_b^t \left[ \frac 1 t \int_0^t x(s) \, ds \right], \label{eq:phi_der}
\end{align}

\noindent which is the average of $x(t)$ with respect to the \emph{backward} distribution over lineages at time $t$, 
\begin{align}
    p_b^t(\ell) &= \frac{ w(t) \, p_f(\ell) }{\E_f[ w(t) ]}. \label{eq:bwd_prob}
\end{align}

\noindent The backward distribution represents genealogies: it is obtained by sampling an individual from the population at time $t$ and tracing back its ancestry \cite{nozoe_inferring_2017}, see Fig.~\ref{fig:intro}. Eq.~\eqref{eq:phi_der}  represents the average value of $x$ over the ancestral history of the current population. This average captures the effect of selection, which favors lineages that result in more offspring; mathematically, it is the spinal decomposition \cite{athreya_branching_1972}. If our population process is ergodic, we can take the thermodynamic limit as $t \rightarrow \infty$ to define the asymptotic backward distribution \cite{ocal_two-clock_2025}. For multitype branching processes, the backward distribution over individuals is the population distribution weighted by the reproductive value \cite{athreya_branching_1972,ocal_two-clock_2025}.

%The appearance of the backward distribution in \eqref{eq:phi_der} reflects the fact that the long-term behavior of a population is reflected in the ancestral history of its individuals. 

Differentiating $\phi$ once more we obtain
\begin{align}
    \phi''(0) &= \lim_{t \rightarrow \infty} \frac 1 t \Var_b^t\left( \int_0^t x(s) \, ds \right), \label{eq:phi_der2}
\end{align}

\noindent which is the (normalized) variance of $x$ over backward lineages. Thus the second derivative yields information about fluctuations in $x$. Similarly, higher-order derivatives of $\phi$ encode higher-order statistics of $x$ measured along backward lineages. 

\section{Selection and phase transitions}

We next study how the fitness of a population changes due to perturbations, which can arise either from genetic mutations or environmental changes. Fitness differences in biology are typically quantified by the selection coefficient $s$ and its infinitesimal version, the selection gradient \cite{gillespie_population_2004,moorad_selection_2013}. As we will see, the selection gradient can be seen as a thermodynamic order parameter and represents a physically measurable quantities that changes discontinuously at a first-order transition.

Assume that our population depends on a control parameter $\beta$, which represents the effect of mutations or environmental changes. If we vary $\beta$ by a small amount $\delta \beta$, the growth rate changes by a corresponding amount $\delta \Lambda$. It is convenient to measure this change by means of the selection coefficient $s$, which is the fractional change in offspring per generation \cite{gillespie_population_2004}. As done in \cite{ilker_modeling_2019,ocal_two-clock_2025}, this can be computed as
\begin{align}
    s &= \frac{e^{(\Lambda + \delta \Lambda) \E_b[\tau]} - e^{\Lambda \E_b[\tau]}}{e^{\Lambda \E_b[\tau]}} \approx (\delta \Lambda) \E_b[\tau], \label{eq:selection_coefficient}
\end{align}

\noindent to first order in $\delta \beta$. Here $\tau$ represents the lifetime of an individual, and we take the typical generation length to be $\E_b[\tau]$ \cite{ocal_two-clock_2025}. The selection coefficient is a dimensionless quantity that predicts e.g.~whether a mutation will outcompete the wild-type population \cite{chevin_measuring_2011} (we discuss genetic drift, which can occur in finite populations, in Apdx.~\ref{apdx:finite}). In our thermodynamic analogy, it represents the free energy difference between two systems. 

Taking the perturbation $\delta \beta$ to be very small, we can define the fitness or selection gradient as
\begin{align}
    \sigma_\beta &:= \left(\frac{\partial \Lambda}{\partial \beta} \right) \E_b[\tau]. \label{eq:selection_differential}
\end{align}

\noindent As the first derivative of a free energy-like function, this has the form of a thermodynamic order parameter, analogous to e.g.~the magnetization in the Ising model. If the control parameter $\beta$ changes the speed at which individuals in the population reproduce (but not the number of offspring), we show in Apdx.~\ref{apdx:el} that 
\begin{align}
    \sigma_\beta &= -\Lambda \E_b\left[ \frac{\partial \tau} {\partial \beta} \right], \label{eq:sigma_formula}
\end{align}

\noindent which is proportional to the infinitesimal change in the generation time of each individual, averaged over the backward distribution \cite{ocal_two-clock_2025}. Eq.~\eqref{eq:sigma_formula} directly follows from the Euler-Lotka formula, which relates the growth rate $\Lambda$ to statistics of backward lineages \cite{pigolotti_generalized_2021,ocal_two-clock_2025}. As a consequence, we remark that $\sigma_\beta$, which determines the response of the population to a perturbation, can be physically measured by looking at the statistics of ancestral lineages, an instance of the fluctuation-dissipation theorem.

Eq.~\eqref{eq:sigma_formula} predicts how a mutation will fare by measuring how it affects the typical lineage in the wild-type population \cite{wakamoto_optimal_2012}. This will be accurate for small mutations that do not significantly change the population make-up. In contrast, we expect \eqref{eq:sigma_formula} to break down if a small change in $\beta$ drastically shifts the population structure, such that lineages that were not prominent in the original population are now selectively favored. Since \eqref{eq:sigma_formula} involves a first derivative of $\Lambda$, this defines a first-order phase transition. Such a first-order transition is witnessed by a discontinuous change in the selection coefficient $\sigma_\beta$, which here plays the role of a latent heat.

\section{Plasmid engineering in bacteria}

\begin{figure} 
    \centering
    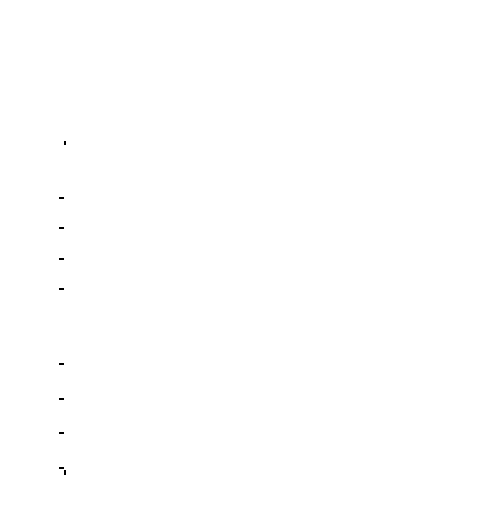
    \caption{\justifying A first-order phase transition in a model of plasmid engineering. The top plot shows the population growth rate $\Lambda$ as a function of the metabolic burden per plasmid $\beta$, for $\alpha = 0.05$. The growth rate is nondifferentiable at a critical value $\beta^* \approx 0.16\%$ (vertical line), beyond which plasmids go extinct. The middle plot shows the selection gradient $\sigma_{\log \beta}$, here taken with respect to $\log \beta$ and estimated using \eqref{eq:sigma_formula}. The bottom plot shows mean plasmid numbers along a backward lineage, which change discontinuously at the critical value: the population can either sustain $k > k^*$ plasmids per ancestral cell, or none at all. Crosses denote Monte Carlo simulations as described in Apdx.~\ref{apdx:simulation} with a population of size $N = 25\, 000$ (averaged over $40$ runs), solid lines the exact solution using \eqref{eq:lambda_exact} and dotted lines the approximation based on \eqref{eq:lambda_approx}. }
    \label{fig:plasmid}
\end{figure}

We demonstrate our ideas on a model of bacterial engineering, where synthetic plasmids are inserted into bacteria to prompt expression of target biomolecules. These plasmids can affect the growth rate of their host cells, in part due to the metabolic burden incurred by the expression of target proteins \cite{glick_metabolic_1995,ilker_modeling_2019}, as well as pathway interference and potential toxic byproducts. In a bioreactor, this results in eventual extinction of plasmids as plasmid-free escape mutants are at a selective advantage. To prevent this, one approach is to render cells ``addicted'' to the plasmid by moving essential bacterial genes onto the latter \cite{rugbjerg_synthetic_2018}; this slows the growth of plasmid-free cells and puts evolutionary pressure on cells to maintain the plasmid. 

We consider a simple model of bacteria such as \textit{Escherichia coli} bearing several copies of a heterologous plasmid that are duplicated and passed onto $m = 2$ daughter cells upon division \cite{paulsson_noise_2001,hernandez-beltran_segregational_2022}. When a cell divides, its plasmids are binomially partitioned between the two daughter cells with 50\% probability each. We assume perfect copy number control, so that each plasmid is duplicated exactly once per generation. Neither binomial partitioning nor perfect copy number control is essential for what follows (see Apdx.~\ref{apdx:nonbinom}). We can describe the number of plasmids $k'$ inherited by a daughter cell using the transition kernel
\begin{align}
    p(k' | k) &= \binom{2k}{k'} \, 2^{-2k}.  \label{eq:plasmid_K}
\end{align}

\noindent In the absence of horizontal gene transfer, lineages cannot regain plasmids once lost, so $k = 0$ is an absorbing state.

We model the fitness effect of plasmids by means of a per-plasmid metabolic burden $\beta$ and an addiction penalty $\alpha$ for cells without plasmids. By a suitable choice of time units we can assume that the time to division of each cell is given by
\begin{align}
    \tau(k) &= \begin{cases} 
        1 + \alpha, & k = 0, \\
        1 + \beta k, & k > 1.
    \end{cases} \label{eq:plasmid_tau}
\end{align}

\noindent The generalized Euler-Lotka equation \cite{pigolotti_generalized_2021} implies that the growth rate $\Lambda$ of this model satisfies
\begin{align}
    \log 2 %&= \lim_{n \rightarrow \infty} -\frac 1 n \log \E_f\left[ e^{-\Lambda \sum_{i=1}^n \tau_i} \right] \\
    &= \Lambda - \lim_{n \rightarrow \infty} \frac 1 n \log \E_f\left[ e^{-\Lambda n \left(\beta \overline{k} + \alpha f_0\right)} \right], \label{eq:el_plasmid}
\end{align}

\noindent where $\overline{k}$ and $f_0$ represent the average number of plasmids and the fraction of plasmid-free cells in a lineage, respectively. Differentiating \eqref{eq:el_plasmid} shows that the fitness gradient \eqref{eq:selection_differential} with respect to $\beta$ equals
\begin{align}
    \sigma_\beta &= - \Lambda \E_b[k], \label{eq:sigma_plasmid}
\end{align}

\noindent which is proportional to the (historic) average plasmid number per cell.

We can understand this model by separately considering the subpopulations consisting of plasmid-bearing and plasmid-free cells. The plasmid-bearing cells reproduce at a rate $0 < \Lambda_+ < \log 2$ given by the implicit equation
\begin{align}
    e^{-\Lambda_+} - e^{-\Lambda_+ (2 + \beta)} &= \frac 1 4, \label{eq:lambda_exact}
\end{align}

\noindent which for small $\beta$ can be approximated by
\begin{align}
    \Lambda_+ &\approx \log 2 - \sqrt{\log(2) \beta}, \label{eq:lambda_approx}
\end{align}

\noindent see Apdx.~\ref{apdx:exact}. In particular, the fitness penalty grows with the square root of $\beta$, which suggests that the number of plasmids per cell should square with the inverse square root. The plasmid-free cells reproduce at a rate
\begin{align}
    \Lambda_0 = \frac{\log 2}{1 + \alpha}.
\end{align} 

\noindent The population supports plasmid-bearing cells in the long run as long as $\Lambda_+ > \Lambda_0$, so that plasmid-free cells do not outcompete plasmid-bearing cells (we discuss finite-size effects due to genetic drift in Apdx.~\ref{apdx:finite}). For fixed $\alpha > 0$, plasmid-bearing cells can persist as long as the metabolic burden $\beta$ is small enough. In this regime, the population growth rate is given by $\Lambda = \Lambda_+$, and combining \eqref{eq:plasmid_tau} with \eqref{eq:sigma_plasmid} yields
\begin{align}
    \E_b[k] &= \frac{1}{2 - e^{\Lambda_+ (1+\beta)}} \approx \frac 1 {2 \sqrt{\log(2) \beta}}. \label{eq:k_mean_exact}
\end{align}

\noindent As the metabolic burden $\beta$ due to plasmids increases, the growth rate $\Lambda_+$ decreases monotonically until we reach the critical point $\beta^*$ satisfying
\begin{align}
    \Lambda_+ \rvert_{\beta = \beta^*} = \Lambda_0, \label{eq:crit_plasmid}
\end{align}

\noindent which using \eqref{eq:lambda_approx} is located around
\begin{align}
    \beta^* &\approx \frac{\log(2) \alpha^2}{(1 + \alpha)^2}. \label{eq:crit_approx}
\end{align}

\noindent If $\beta$ increases beyond this value, plasmid-bearing cells grow at a slower rate than plasmid-free cells, and the latter take over the population \textemdash{} in this regime, $\Lambda = \Lambda_0$ and $\E_b[k] = 0$. The point $\beta^*$ represents a first-order phase transition in \eqref{eq:sigma_formula}, where the average number of plasmids per ancestral cell jumps to $0$ from a threshold value 
\begin{align}
    k^* &\approx \frac{1 + \alpha}{2 \log(2) \alpha}. \label{eq:k_crit_approx}
\end{align}

\noindent Here we combine our approximations \eqref{eq:k_mean_exact} and \eqref{eq:crit_approx}. Thus $k^*$ represents the minimal number of plasmids (per ancestral cell) that can be stably maintained in a population, which corresponds to the latent heat of the phase transmission. We illustrate this model in Fig.~\ref{fig:plasmid} using numerical Monte Carlo simulations based on \cite{giardina_direct_2006,lecomte_numerical_2007,nemoto_finite-time_2017}, as well as the exact solution \eqref{eq:lambda_exact}; for $\alpha = 5\%$, the threshold value is roughly $k^* \approx 15$. We provide a more detailed mathematical analysis of this model in Apdx.~\ref{apdx:model} based on a matrix representation \cite{bapat_nonnegative_1997,kato_perturbation_1995}. 

Our model explicitly predicts a first-order transition in bacterial population dynamics. One can see directly that this phase transition must necessarily be discontinuous: since plasmid-free cells can only produce plasmid-free offspring, the ancestral lineage of a plasmid-containing cell must maintain $k \geq 1$ plasmids throughout (this also follows from \eqref{eq:k_mean_exact}). In contrast, the ancestral lineage of a plasmid-free cell will eventually contain no plasmids, which implies that its long-term average must be $0$. We hypothesize that an extension of this model to incorporate horizontal plasmid transmission \cite{rodriguez-beltran_beyond_2021} could exhibit higher-order transitions by allowing plasmids to be regained in lineages.

\section{Discussion}

In this paper we demonstrate the existence of phase transitions in microbial populations growth, under metabolic constraint, using a large deviation formalism along the same lines as in classical statistical physics \cite{touchette_large_2009,sughiyama_pathwise_2015,genthon_fluctuation_2020,levien_large_2020,ocal_two-clock_2025}. We connect singularities in the population growth rate $\Lambda$ to sudden changes in population structure as measured by the selection coefficient $s$ in population genetics, which acts as a thermodynamic order parameter. In contrast to classical treatments, our approach is mesoscopic in nature and works on the level of lineages, where genuinely new phenomena emerge \cite{anderson_more_1972}. In particular, we show that fitness gradients, an important concept in evolutionary biology, represent physically measurable quantities. We illustrate our results with an industrially relevant model of bacterial plasmid transmission.

Our approach complements existing work that identifies phase transitions in biological populations \cite{munoz_colloquium_2018,ordway_phase_2020,heffern_phase_2021,scott_phase_2022}. Surprisingly, we find that environmental fluctuations are not necessary to induce sharp transitions in population-level behavior as in \cite{skanata_evolutionary_2016,mayer_transitions_2017}: in our example model, the intrinsic stochasticity of plasmid inheritance is sufficient. We expect that tractable models similar to the example studied in this paper will both provide biological insight and contribute interesting examples of physical phase transitions. %Our mathematical framework can provide a new context in which phase transitions such as () can be studied.

\subsection*{Data availability}
\noindent The code used for numerical simulations in this paper is accessible at \url{https://github.com/kaandocal/microbialphasetransitions}.

\subsection*{Acknowledgments}
\noindent The authors would like to thank Augustinas Sukys and the anonymous referees for their feedback, and gratefully acknowledge financial support through an ARC Laureate Fellowship to MPHS (FL220100005).

\subsection*{Declaration of Interests}
\noindent MPHS is co-founder, shareholder, director, and CSO of Cell Bauhaus Pty Ltd.

\bibliography{references}

\clearpage 

\onecolumngrid

\appendix

\setcounter{page}{1}

\renewcommand\thefigure{S\arabic{figure}}   
\setcounter{figure}{0}
%TC:ignore
\section{Partition function}

\label{apdx:cgf}

We justify the scaling in the partition function \eqref{eq:pf_def} by noting that $X(t)$ scales with $t$. Eq.~\eqref{eq:phi_der} follows directly from
\begin{align}
    \phi'(\lambda) &= \lim_{t \rightarrow \infty} \frac 1 t \frac{\E_f[ w(t) X(t) e^{\lambda X(t)} ]}{\E_f[ w(t) e^{\lambda X(t)} ]}.
\end{align}

\noindent Differentiating this at $\lambda = 0$, we obtain
% \begin{align}
%     \phi''(\lambda) &= \lim_{t \rightarrow \infty} \frac 1 t \left(\frac{\E_f[ w(t) X(t)^2 e^{\lambda X(t)} ]}{\E_f[ w(t) e^{\lambda X(t)}]} - \frac{\E_f[ w(t) X(t) e^{\lambda X(t)} ]^2}{\E_f[ w(t) e^{\lambda X(t)}]^2}\right), 
% \end{align}

% \noindent and so
\begin{align}
    \phi''(0) &= \lim_{t \rightarrow \infty} \frac 1 t \left( \frac{\E_f[ w(t) X(t)^2]}{\E_f[ w(t)]} - \frac{\E_f[ w(t) X(t)]^2}{\E_f[ w(t) ]^2} \right) = \lim_{t \rightarrow \infty} \frac 1 t \left( \E_b^t[ X(t)^2 ] - \E_b^t[ X(t)]^2\right),
\end{align}

\noindent  which proves \eqref{eq:phi_der2}.

\section{Computing growth rates}

\label{apdx:el}

The definition of the growth rate $\Lambda$ in \eqref{eq:lambda_def} involves the counting process $w(t)$ and is not always suitable for direct computation. Based on \cite{pigolotti_generalized_2021,ocal_two-clock_2025} we derive an alternative formula by a change of variable from physical time $t$ to generation $n$. In analogy to \eqref{eq:pop_abs_fitness} we define the weight of a lineage in generation $n$ as 
\begin{align}
    w_n &= \prod_{i=1}^n m_i. \label{eq:pop_abs_fitness_N}
\end{align}

\noindent Given a time-dependent quantity $x(t)$ we let
\begin{align}
    X_n &:= X(t_n) = \int_0^{t_n} x(s) \, ds.
\end{align}

\noindent We define the forward distribution on lineages in generation $n$ exactly as in the main text. This defines a different thermodynamic ensemble over lineages, the $n$-ensemble, whose partition function is defined as 
\begin{align}
    \psi(\lambda) &= \lim_{n \rightarrow \infty} \frac 1 n \log \E_f \left[ w_n e^{-\Lambda t_n + \lambda X_n} \right]. \label{eq:psi_def}
\end{align}

\noindent Note the extra term $e^{-\Lambda t_n}$. We justify this definition in Apdx.~\ref{apdx:duality_proof} using the approach in \cite{ocal_two-clock_2025}. Briefly, we define the extended partition functions
\begin{align}
    \tilde \phi(\xi, \lambda) &= \lim_{t \rightarrow \infty} \frac 1 t \log \E_f[ w(t) \, e^{-\xi n(t) + \lambda X(t)}],  \label{eq:phi_aug}\\
    \tilde \psi(\alpha, \lambda) &= \lim_{n \rightarrow \infty} \frac 1 n \log \E_f \left[ w_n e^{-\alpha t_n + \lambda X_n} \right], \label{eq:psi_aug}
\end{align}

\noindent which are shown in Apdx.~\ref{apdx:duality_proof} to satisfy
\begin{align}
    \tilde \phi(\tilde \psi(\alpha, \lambda), \lambda) &= \alpha. \label{eq:psi_duality}
\end{align}

\noindent By repeatedly differentiating this identity with respect to $\alpha$ and $\lambda$, we can express thermodynamic observables such as \eqref{eq:phi_der} and \eqref{eq:phi_der2} in terms of $\psi$, e.g.
\begin{align}
    \psi'(0) &= \frac 1 {\E_b[\tau]} \phi'(0),
\end{align}

\noindent where $\E_b[\tau]$ is the mean generation time over backward lineages. Accordingly, we interpret
\begin{align}
    p_b^n(\ell) &= \frac{ w_n(\ell) \, e^{-\Lambda t_n} p_f(\ell) }{\E_f[ w_n(\ell) \, e^{-\Lambda t_n} ]} \label{eq:bwd_prob_N}
\end{align}

\noindent as the backward distribution over lineages when $n$ is fixed instead of $t$, in analogy to \eqref{eq:bwd_prob} \cite{ocal_two-clock_2025}. The two backward distributions do not agree for finite $t$ and $n$, but in the limits as $t \rightarrow \infty$ and $n \rightarrow \infty$ they yield the same thermodynamic behavior due to \eqref{eq:psi_duality}.

As a direct consequence of \eqref{eq:psi_duality} we obtain the generalized Euler-Lotka equation of \cite{pigolotti_generalized_2021,ocal_two-clock_2025}:
\begin{align}
    \lim_{n \rightarrow \infty} \frac 1 n \log \E_f\left[ \left(\prod_{i=1}^n m_i \right) e^{-\Lambda \sum_{i=1}^n \tau_i} \right] &= 0. \label{eq:el}
\end{align}

\noindent In our bacterial example where every cell divides in two, $m_i \equiv 2$ and \eqref{eq:el} simplifies to the main result in \cite{pigolotti_generalized_2021}:
\begin{align}
    \lim_{n \rightarrow \infty} \frac 1 n \log \E_f\left[ e^{-\Lambda \sum_{i=1}^n \tau_i} \right] &= -\log 2. \label{eq:el_2}
\end{align}

Following \cite{ocal_two-clock_2025}, we now introduce slight perturbations of the form
\begin{align}
    \tau_i &\mapsto \tau_i + \delta \tau_i, & \Lambda \mapsto \Lambda + \delta \Lambda,
\end{align}

\noindent while keeping $m_i$ the same; biologically we interpret this as a mutation. To first order in the perturbation we can then write
\begin{align}
    \frac 1 n \log \E_f\left[ w_n e^{-(\Lambda + \delta \Lambda) \sum_{i=1}^n (\tau_i + \delta \tau_i)} \right] &\approx \frac 1 n \log \E_f\left[ w_n e^{-\Lambda \sum_{i=1}^n \tau_i} \right] - (\delta \Lambda) \frac 1 n \sum_{i=1}^n \frac{\E_f\left[ w_n e^{-\Lambda \sum_{i=1}^n \tau_i} \tau_i \right]}{\E_f\left[ w_n e^{-\Lambda \sum_{i=1}^n \tau_i} \right]} \nonumber \\
    &\quad - \Lambda \frac 1 n \sum_{i=1}^n \frac{\E_f\left[ w_n e^{-\Lambda \sum_{i=1}^n \tau_i} (\delta \tau_i) \right]}{\E_f\left[ w_n e^{-\Lambda \sum_{i=1}^n \tau_i} \right]}.
\end{align}

\noindent In the limit as $n \rightarrow \infty$, the Euler-Lotka equation \eqref{eq:el_2} applied to the original and the perturbed population implies that the left-hand side and the first term on the right-hand side both converge to $0$. In view of \eqref{eq:bwd_prob_N}, we can interpret the remaining two terms as expectations over the backward lineage distribution in generation $n$. As a result,
\begin{align}
    0 &= (\delta \Lambda) \lim_{n \rightarrow \infty} \frac 1 n \sum_{i=1}^n \E_b^n[ \tau_i ] + \Lambda \lim_{n \rightarrow \infty} \frac 1 n \sum_{i=1}^n \E_b^n[ \delta \tau_i ].
\end{align}

\noindent This combined with \eqref{eq:selection_coefficient} directly yields \eqref{eq:selection_differential} in the limit.

\section{Analysis of the plasmid model}

\label{apdx:model}

\begin{figure}
    \centering
    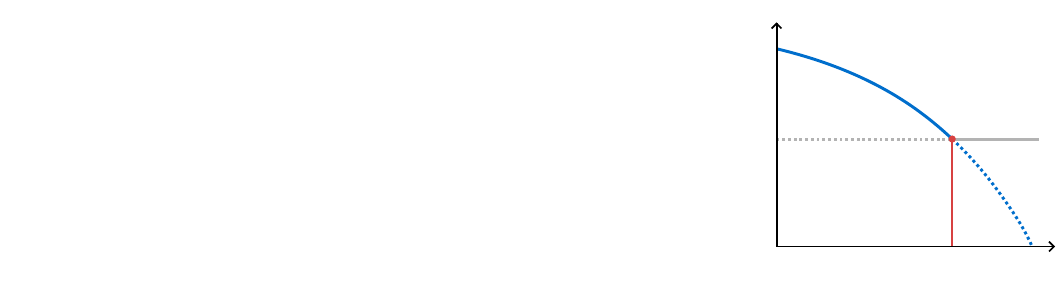
    \caption{\justifying \textbf{a.}~Heat map of the growth rate $\Lambda$ for the plasmid model obtained using \eqref{eq:lambda_exact} The phase transition manifests as a kink in the growth rate. \textbf{b.}~Mean plasmid count along backward lineages for the same parameter combinations. The phase transition is visible as a sudden drop from a threshold value $k^* > 0$ to $0$ as $\beta$ increases past the critical threshold, determined by $\alpha$. \textbf{c.}~Diagram of the phase transition for fixed $\alpha$. The plasmid-bearing and plasmid-free subpopulations can be represented by their steady-state distributions, which are eigenvectors of a certain matrix $M(\nu)$ depending on $\alpha$ and $\beta$. The two subpopulations grow at different rates that are encoded by their respective eigenvalues. As long as they do not intersect, these eigenvalues and eigenvectors are smooth functions of $\alpha$ and $\beta$. At a critical value $\beta^*$, the two eigenvalues coalesce and the two subpopulations grow at the same asymptotic rate, forming a two-dimensional space of eigenvectors. Upon crossing $\beta^*$, the dominant subpopulation changes suddenly: the growth rate (related to the dominant eigenvalue) is nondifferentiable at that point, and the backward and population distributions (related to the corresponding eigenvectors) exhibit a jump discontinuity.}
    \label{fig:plasmid_apdx}
\end{figure}

We can represent the plasmid model as an age-dependent multitype branching process, where each cell is defined by the number of plasmids $k$ inherited at birth and the lifetime of a cell is given by \eqref{eq:plasmid_tau}. Most of the results below are standard in branching process theory \cite{athreya_branching_1972} and are discussed in \cite{ocal_two-clock_2025}. The idea is that the growth rate $\Lambda$ is encoded in the eigenvalues of a certain matrix associated with the branching process, and a careful analysis of these eigenvalues reveals that $\Lambda$ is nondifferentiable at points where two eigenvalues coalesce, with a sudden jump in the corresponding eigenvectors which represent the backward distribution (Fig.~\ref{fig:plasmid}c). This simple observation is somewhat obscured by the fact that $\Lambda$ is defined implicitly in terms of the eigenvalues.

For a given $\nu > 0$ we define the time-discounted next-generation matrix
\begin{align}
    M(\nu)_{k'k} &= 2 e^{-\nu \tau(k)} \, p(k' \, | k).
\end{align}

\noindent Then $M(0)$ is the ordinary next-generation matrix. $M(\nu)$ is infinite-dimensional as the number of plasmids is theoretically unbounded, but this is not essential to our argument.

Let $\pi(k, t)$ be the number of newborn cells at time $t$. In the long-time limit when the population is at steady state, this function becomes stationary in the sense that
\begin{align}
    \pi(k, t) &= \pi_\infty(k) \, N(t), \label{eq:pop_p_t}
\end{align}

\noindent where $\pi_\infty$ necessarily satisfies 
\begin{align}
    \sum_{k=0}^\infty \pi_\infty(k) &= 1. \label{eq:pop_p_inf}
\end{align}

\noindent The vector $\pi_\infty$ represents the steady-state distribution of newborn cells in the population. We can relate the average number of newborn cells at time $t$ to the cells born in the past as
\begin{align}
    \pi(k', t) &= 2 \sum_{k=0}^\infty  p(k' | k) \pi(k, t - \tau(k)).
\end{align}

\noindent Since $N(t) \sim e^{\Lambda t}$ at steady state, combining this with \eqref{eq:pop_p_t} and \eqref{eq:pop_p_inf} yields
\begin{align}
    \pi_\infty(k') &= \sum_{k=0}^\infty 2 p(k' | k) e^{-\Lambda \tau(k)} \pi_\infty(k) = \sum_{k=0}^\infty M(\Lambda)_{k',k} \pi_\infty(k), \label{eq:pi_M}
\end{align}

\noindent so that $\pi_\infty$ is a right eigenvector of $M(\Lambda)$ with eigenvalue $1$. Conversely, every such eigenvector represents a steady state distribution of the population that grows with rate $\Lambda$. If $h$ is the corresponding left eigenvector, one can show that the steady-state distribution of cells in the backward distribution equals \cite{ocal_two-clock_2025}
\begin{align}
    \pi_b(k) &\propto \pi_\infty(k) h(k). \label{eq:pi_b_p}
\end{align}

\noindent This shows that we can characterize the long-term growth rate $\Lambda$ and the backward distribution over cells in terms of the eigenvalues of $M(\nu)$.

The matrix $M(\nu)$ has nonnegative entries that depend smoothly on $\alpha, \beta$ and $\nu$. By the Perron-Frobenius theorem \cite{bapat_nonnegative_1997}, its dominant eigenvalue is real and positive, and has associated left and right eigenvectors $v$ and $w$ with nonnegative entries. The dominant eigenvalue is equal to the spectral radius $\rho$ of $M(\nu)$, which is a strictly increasing, log-convex function of $\nu$. By the discussion above, $\Lambda$ is characterized by the requirement that its dominant eigenvalue is $1$.

The growth rate $\Lambda$ is defined by the equation
\begin{align}
    \rho(M(\Lambda; \alpha, \beta)) &= 1,
\end{align}

\noindent where $\rho$ is the spectral radius; this shows that $\Lambda$ is a continuous function of $\alpha$ and $\beta$. As long as the dominant eigenvalue of $M(\Lambda)$ has multiplicity one, the corresponding left and right eigenvectors $v(\alpha,\beta)$ and $w(\alpha, \beta)$ trace out smooth curves as functions of $\alpha$ and $\beta$ \cite{kato_perturbation_1995}, see Fig.~\ref{fig:plasmid_apdx}c. When the dominant eigenvalue crosses another eigenvalue, the eigenspace becomes two-dimensional and the dominant eigenvector is no longer uniquely defined. At this point, the latter can jump discontinuously, and the dominant eigenvalue $\Lambda$ may not be differentiable.

We now observe that the basis vector $e_0$ corresponding to the state $k = 0$ is always a right eigenvalue of $M(\Lambda)$,
\begin{align}
    M(\Lambda) e_0 &= 2 e^{-\Lambda (1 + \alpha)} e_0,
\end{align}

\noindent since a cell with $0$ plasmids can only give rise to offspring with $0$ plasmids. Since the largest eigenvalue of $M(\Lambda)$ is $1$, this is the dominant eigenvector precisely when
\begin{align}
    \Lambda &= \frac {\log 2}{1 + \alpha}, \label{eq:stable_boundary}
\end{align}

\noindent cf.~\eqref{eq:crit_plasmid}. In this case the population distribution $\pi_\infty$ contains only plasmid-free cells, and by \eqref{eq:pi_b_p} the same holds for the backward distribution. This represents the regime where plasmid-free cells are at a selective advantage and plasmids eventually become extinct from the population.

We can now understand our phase transition in more detail. When the metabolic burden $\beta$ is low enough, the population of plasmid-bearing cells will grow faster than the plasmid-free cells, and we have
\begin{align}
    \Lambda &> \frac{\log 2}{1 + \alpha}. \label{eq:stable_regime}
\end{align}

\noindent As long as this holds, the population distribution $\pi_\infty$ and the backward distribution $\pi_b$ over plasmid numbers change continuously with $\alpha$ and $\beta$. This requires that $\beta$ be less than the threshold $\beta^*$ defined by \eqref{eq:crit_plasmid}. At the threshold, the eigenvalue corresponding to the stable plasmid distribution crosses the eigenvalue for the plasmid-free population. As described above, the space of eigenvectors becomes two-dimensional, and for $\beta > \beta^*$ the population distribution $\pi_\infty$ and the backward distribution $\pi_b$ jump to $e_0$, which corresponds to the sudden extinction of plasmid-bearing cells. 

The growth rate in the regime given by \eqref{eq:stable_regime} is independent of $\alpha$, since plasmid-free cells grow at a strictly slower rate than plasmid-bearing cells. In the extinction regime, the opposite occurs and $\Lambda$ is independent of $\beta$. As a result, this model has a very simple structure that we visualize in the phase diagram plotted in Fig.~\ref{fig:plasmid_apdx}: to the left of the phase boundary, the growth rate and mean plasmid concentration only depend on $\beta$, to the right they only depend on $\alpha$ and the mean plasmid concentration is $0$. The location of the phase boundary is obtained by solving \eqref{eq:crit_plasmid}.

\section{Derivation of the exact growth rate}

\label{apdx:exact}

In this section we derive Eq.~\eqref{eq:lambda_exact}. We restrict our attention to the plasmid-bearing cells only, and modify our model so that plasmid-free cells are removed at birth. We can achieve this by setting $M(\nu)_{0k} = 0$ for $k \geq 1$, which implies that $\pi_\infty(0) = 0$. Introduce the cumulant generating function
\begin{align}
    \phi(z) &= \sum_{k=1}^\infty \pi_\infty(k) z^k, \label{eq:pgf_phi_def}
\end{align}

\noindent which is analytic for $\left|z\right| < 1$. The recurrence relation \eqref{eq:pi_M} with our modification of $M$ yields the functional equation
\begin{align}
    \phi(z) &= 2e^{-\Lambda_+} \sum_{k'=0} z^{k'} \sum_{k=0}^\infty p(k' | k) e^{-\Lambda_+ \beta k} \pi_\infty(k) - 2 e^{-\Lambda_+} \sum_{k=0}^\infty p(0 | k) e^{-\Lambda_+ \beta k} \pi_\infty(k) \\
    &= 2 e^{-\Lambda_+} \phi\left(\frac{(1 + z)^2}{4 e^{\Lambda_+ \beta}}\right) - 2 e^{-\Lambda_+} \phi\left(\frac 1 {4 e^{-\Lambda_+ \beta}}\right), \label{eq:lambda_exact_phi_formula}
\end{align}

\noindent where the second term arises since we remove cells with $0$ plasmids. The second equality follows from the fact that the transition kernel $p(k' | k)$ amounts to doubling, followed by binomial subsampling with probability $1/2$.

Eq.~\eqref{eq:lambda_exact_phi_formula} suggests that we study $\phi(z)$ by looking at fixed points of the map 
\begin{align}
    z \mapsto \frac{(1 + z)^2}{4e^{\Lambda_+ \beta}}. \label{eq:pgf_T_map}
\end{align}

\noindent There are two fixed points, given by
\begin{align}
    z_{\pm} &= 2 e^{\Lambda_+ \beta} - 1 \pm \sqrt{\left(2 e^{\Lambda_+ \beta} - 1\right)^2 - 1}.
\end{align}

\noindent Assuming that $\Lambda_+ > 0$, $z_-$ is a fixed point with $0 < z_- < 1$. Since $\left| z_- \right| < 1$, $\phi$ is analytic around $z_-$, and differentiating Eq.~\eqref{eq:lambda_exact_phi_formula} at $z = z_-$ yields 
\begin{align}
    \phi'(z_-) &= e^{-\Lambda_+(1 + \beta)} \left(1 + z_-\right) \phi'(z_-).
\end{align}

\noindent Now \eqref{eq:pgf_phi_def} together with $z_- > 0$ imply that $\phi'(z_-) > 0$, so we can cancel the derivative on both sides to obtain
\begin{align}
    z_- &= e^{\Lambda_+(1 + \beta)} - 1.
\end{align}

\noindent As a result, the right-hand side must be a fixed point of the map \eqref{eq:pgf_T_map}; equating the two sides and simplifying yields \eqref{eq:lambda_exact}. 

Since $\tau = 1 + \beta k$, the fluctuation relation \eqref{eq:sigma_formula} reads
\begin{align}
    \frac{\partial \Lambda}{\partial \beta} \left(1 + \beta \E_b[k]\right) &= -\Lambda \E_b[k]. \label{eq:apdx_plasmid_fr}
\end{align}

\noindent Implicitly differentiating \eqref{eq:lambda_exact} with respect to $\beta$ we obtain
\begin{align}
    \frac{\partial \Lambda_+}{\partial \beta} &= - \frac {\Lambda_+}{2 + \beta - e^{\Lambda_+(1 + \beta)}},
\end{align}

\noindent and combining these two identities yields Eq.~\eqref{eq:k_mean_exact}.

For small $\beta$, $\Lambda_+ \approx \log 2$ and we can attempt an expansion of the form
\begin{align}
    \Lambda_+ &\approx \log 2 - c \beta^q + o(\beta^q) \label{eq:apdx_lambda_ansatz}
\end{align}

\noindent around $\beta = 0$ for some $q > 0$. Plugging this into \eqref{eq:lambda_exact} and expanding the lowest-order terms yields
\begin{align}
    \frac 1 2 e^{-c \beta^q} - \frac 1 4 e^{-2 c \beta^q + c \beta^{q+1} - \beta \log 2} &= \frac 1 4.
\end{align}

\noindent Expanding to first order in $\beta^q$ suggests that $q < 1$, since otherwise the lowest-order terms simplify to $\beta \log 2 = 0$. Expanding to second order in $\beta^q$ and to first order in $\beta$ we obtain
\begin{align}
    \frac 1 2 \left(1 - c \beta^q + \frac {c^2} 2 \beta^{2q}\right) - \frac 1 4 \left(1-2c\beta^q+2c^2 \beta^{2q} - \beta \log 2\right) &= \frac 1 4.
\end{align}

\noindent The ansatz $2q = 1$ then yields our approximation \eqref{eq:lambda_approx}. The approximation in \eqref{eq:k_mean_exact} follows by Taylor expanding the exponential to first order.

We can extend the above approach to any model where plasmids reproduce independently, so that we can write
\begin{align}
    k' &= \sum_{i=1}^k X_i,
\end{align}

\noindent with $X_i \in \{0, 1, 2, \ldots\}$ iid.~variables that represent the number of copies of the $i$-th plasmid passed on to the daughter cell. In the above example, $X_i \sim \mathrm{Binom}(2, 1/2)$. If $\psi(z)$ is the probability generating function of the $X_i$, then a similar argument to the above leads us to the equation
\begin{align}
    \phi(z) &= 2e^{-\Lambda_+} \phi(\psi(z) e^{-\Lambda_+ \beta}) - 2 e^{-\Lambda_+} \phi(\psi(0) e^{-\Lambda_+ \beta}). \label{eq:lambda_exact_phi_psi_formula}
\end{align}

\noindent We assume that $\psi'(1) = 1$, ie.~$\E[X_i] = 1$, and let 
\begin{align}
    \sigma^2 = \psi''(1) &= \Var(X_i).
\end{align}

\noindent The case $\sigma = 0$ represents perfect plasmid transmission, in which case there cannot be a phase transition, so we assume $\sigma > 0$. In particular, for any $\Lambda_+ > 0$ there exists a unique $z_* \in (0, 1)$ satisfying
\begin{align}
    z_* &= \psi(z_*) e^{-\Lambda_+ \beta}. \label{eq:apdx_psi_fixed_point}
\end{align}

\noindent Differentiating \eqref{eq:lambda_exact_phi_psi_formula} at $z_*$ yields the compatibility condition
\begin{align}
    1 &= 2e^{-\Lambda_+ (1 + \beta)} \phi'(z_*).  \label{eq:apdx_psi_fp}
\end{align}

\noindent We now assume that $\beta$ is small, make the ansatz in \eqref{eq:apdx_lambda_ansatz} with $q = 1/2$ and write $z_* = 1 - \delta$ for $\delta$ small. Taylor expanding \eqref{eq:apdx_psi_fixed_point} to first order in $\beta$ and second order in $\beta$ yields
\begin{align}
    1 - \delta &\approx \left(1 - \delta - \frac{\sigma^2}{2} \delta^2\right) e^{-\beta \log 2},
\end{align}

\noindent from which it follows that
\begin{align}
    \delta \approx \frac{\sqrt{2 \log(2) \beta}}{\sigma}.
\end{align}

\noindent The compatibility condition \eqref{eq:apdx_psi_fp} then reads
\begin{align}
    1 &\approx e^{c\sqrt{\beta}} (1 - \sigma^2 \delta) \approx e^{c \sqrt{\beta}} \left(1 - \sigma \sqrt{2 \log(2)\beta}\right),
\end{align}

\noindent and by comparing terms of order $\sqrt{\beta}$ we arrive at
\begin{align}
    \Lambda_+ &\approx \log 2 - \sigma \sqrt{2 \log(2)\beta}.
\end{align}

\noindent The fluctuation relation \eqref{eq:apdx_plasmid_fr} additionally implies that
\begin{align}
    \E_b[k] &\approx \frac{\sigma}{\sqrt{2 \log(2) \beta}}.
\end{align}

\noindent In Apdx.~\ref{apdx:nonbinom}, we apply this to the Poisson model with $\psi(z) = e^{z-1}$ and the negative binomial model with $\psi(z) = 1/(2-z)$.

\section{Extensions of the plasmid model}

\label{apdx:nonbinom}

Our model of plasmid transmission assumes perfect copy number control (each plasmid is duplicated exactly once per generation) together with binomial (independent) partitioning, which is encapsulated in \eqref{eq:plasmid_K}. In this section we verify that our results do not depend on these assumptions. Fundamentally, the phase transition between survival and extinction is caused by the opposing forces of selection (due to metabolic load and addiction) and genetic drift (due to random segregation). If plasmid inheritance is highly stochastic, extinction due to genetic drift becomes more likely, and we expect a phase transition to occur for lower values of $\beta$ and higher values of $\E_b[k]$. In contrast, if inheritance is mostly deterministic, then low plasmids numbers can be sustained for longer in the population and we expect a phase transition at higher $\beta$ and lower $\E_b[k]$.

We first consider two alternative inheritance mechanisms, still assuming perfect copy number control. In the first, plasmids inherited by a daughter cell follow a (symmetric) beta-binomial distribution with parameter $w$:
\begin{align}
    p_2(k' | k) &= \binom{2k}{k'} \frac{B(k' + w, 2k - k' + w)}{B(w,w)}, \label{apdx:betabinom}
\end{align}

\noindent where 
\begin{align}
    B(a, b) &= \frac{\Gamma(a) \Gamma(b)}{\Gamma(a + b)}
\end{align}

\noindent is the Beta function. This can be seen as a model of asymmetric cell division where each daughter cell inherits a random fraction $f \sim B(w, w)$ of its parent's contents, and plasmids are binomially partitioned between the daughter cells in proportion to their size \cite{ocal_cell_2025}. The variability in $f$ increases segregation noise compared to the binomial model, where $f \equiv 1/2$ deterministically. Another source of segregation noise is clustering of plasmids (multimerization) \cite{paulsson_noise_2001}.

Our second model assumes that each daughter cell inherits exactly $k$ plasmids with probability $1 - 2q$, and gains or loses one plasmid with probability $q$ each, where $0 \leq q \leq 1/2$. Plasmid-free cells do not transmit any plasmids, so that
\begin{align}
    p_3(k'|k > 0) &= \begin{cases} 1 - 2q, & k' = k, \\ 
        q, & k' = k \pm 1,
    \end{cases} & p_3(k'|0) &= \delta(k'),  \label{apdx:randomwalk}
\end{align}

\noindent This can be seen as a simple random walk with absorbing boundary at $0$. Since plasmid numbers can change by at most $1$ between generations, this model features less segregation noise compared to the binomial model. 

We now consider the case of inexact copy number control \cite{paulsson_noise_2001}. We first assume that the number of replicates of each plasmid is Poisson distributed with mean $2$, so under binomial partitioning of the plasmids between daughter cells the transition kernel is given by a Poisson distribution with mean $k$:
\begin{align}
    p_4(k' | k) &= e^{-k} \frac{k^{k'}}{k'!}. \label{eq:apdx_cnc_poisson}
\end{align}

\noindent This is the case if replication events for each plasmid are Markovian, as long as replicated plasmids cannot be replicated again within one generation. If copied plasmids can themselves be replicated within a generation, we can model plasmid numbers as an exponentially growing Yule process, in which case the number of plasmids upon division is negative binomially distributed. Again assuming that each plasmid is on average doubled in a generation, under binomial partitioning we obtain a negative binomial transition kernel with shape $k$ and mean $k$:
\begin{align}
    p_5(k' | k) &= \binom{k+k'-1}{k'} 2^{-k-k'}.\label{eq:apdx_cnc_negbin}
\end{align}

\begin{figure}
    \centering
    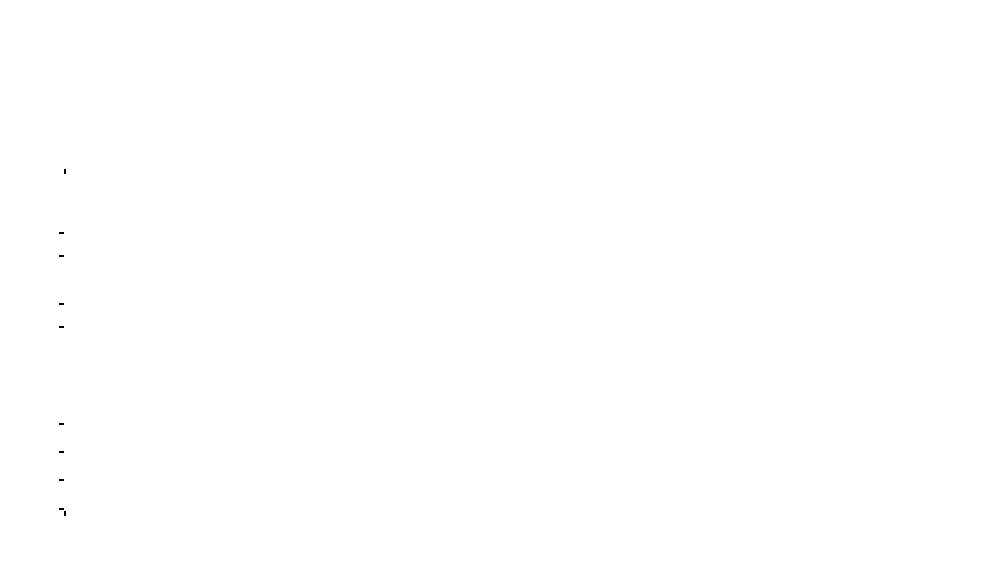
    \caption{Different models of plasmid transmission predict the same phase transition. \textbf{a.} Non-binomial models of plasmid partitioning. Crosses indicate the binomial model in Fig.~\ref{fig:plasmid}, circles the beta-binomial model \eqref{apdx:betabinom} with $w = 10$, and triangles the random walk model \eqref{apdx:randomwalk} with $q = 1/4$. High variability during partitioning decreases the critical value $\beta^*$ and increases the minimal stable plasmid number $k^*$. \textbf{b.} Different models of copy number control. Crosses indicate exact copy number control in Fig.~\ref{fig:plasmid}, circles the Poisson model in \eqref{eq:apdx_cnc_poisson}, and triangles the negative binomial model in \eqref{eq:apdx_cnc_negbin}. Data were obtained using Monte Carlo simulations as described in Apdx.~\ref{apdx:simulation} with a population size $N = 25\,000$, averaged over $40$ runs. Dotted lines indicate the approximation \eqref{eq:apdx_approx_cnc}.}
    \label{fig:plasmid_nonbinom}
\end{figure}

\begin{figure}
    \centering
    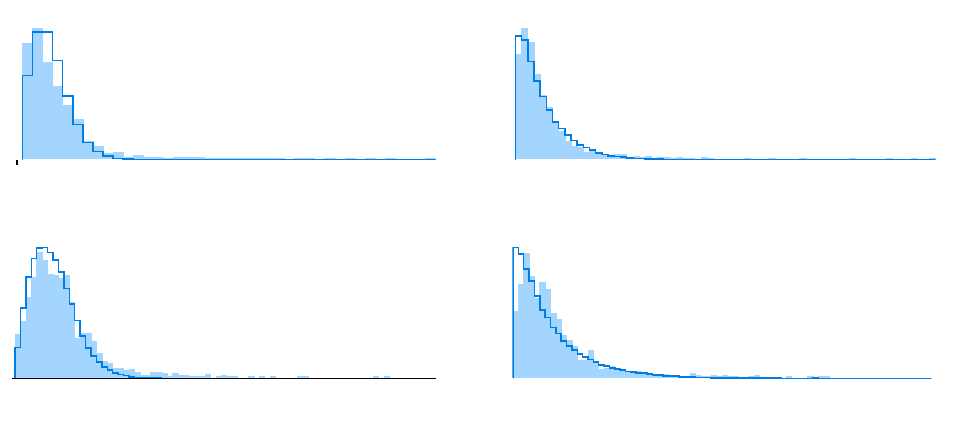
    \caption{Predicted vs.~measured plasmid copy numbers using data obtained from \cite{shao_single-cell_2021} for four different plasmids. The plots display steady-state population distributions for plasmid-bearing cells only. We assume perfect copy number control and fit $\beta$ and the partitioning model separately for each plasmid. While ColE1 was not reported to have an active partitioning mechanism, using binomial partitioning led to overly wide distributions. Predicted distributions were obtained by simulating a fixed population of size $N = 10\,000$ to steady state and averaging over $8$ runs.}
    \label{fig:plasmid_data}
\end{figure}

\noindent As we show in Apdx.~\ref{apdx:exact}, these models behave similarly to our original model, and for small $\beta$
\begin{align}
    \Lambda_+ &\approx \log 2 - \sigma \sqrt{2 \log(2) \beta}, &\E_b[k] &\approx \frac {\sigma} {\sqrt{2 \log(2) \beta}}, \label{eq:apdx_approx_cnc}
\end{align}

\noindent with $\sigma = 1$ in the Poisson case and $\sigma = \sqrt{2}$ in the negative binomial case. The behavior of the models discussed in this section is illustrated in Fig.~\ref{fig:plasmid_nonbinom}, which shows the same discontinuous behavior in all cases. We compare the plasmid number distributions predicted by these models to experimental data from \cite{shao_single-cell_2021} in Fig.~\ref{fig:plasmid_data}. We obtain good fits to the bulk of the data, although the experimental measurements include a small fraction of outlier cells with very high plasmid numbers that were not well captured by our models. 

\section{Finite populations}

\label{apdx:finite}

Our derivations assume exponential population growth, but real populations cannot grow indefinitely. In finite populations, genetic drift can lead to extinction of a beneficial mutant, or fixation of a deleterious mutant. Here we consider the behaviour of populations with a large, but finite carrying capacity $N$. Every time a cell divides in two, resulting in $N + 1$ cells, one cell from the population is removed at random; this represents a generalization of the Moran process. The probability of fixation can be approximated using Kimura's formula \cite{kimura_probability_1962} as a function of $s$ and the population size $N$:
\begin{align}
    p_{\mathrm{fix}} &= \frac{1 - e^{-Nsf}}{1 - e^{-Ns}},
\end{align}

\noindent where $f$ is the initial fraction of mutants. This suggests that the probability of extinction or fixation against the selection gradient, which is essentially a large-deviation event, decreases exponentially in $N$. This is similar to the behavior observed in many thermodynamic systems, where macroscopic fluctuations become exponentially rare in the system size. Similar scaling arguments underlie the use of cloning algorithms \cite{giardina_direct_2006,lecomte_numerical_2007}, which simulate branching processes with a fixed carrying capacity to estimate rate functions in nonequilibrium systems.

Our plasmid model suffers from a more serious finite-size effect. Since plasmid-bearing cells can give rise to plasmid-free cells, but not vice versa, there is a unique absorbing state where all cells have lost their plasmids. As a result, the probability that plasmids will eventually die out in a finite population is $1$, even if plasmid-bearing cells grow faster: the presence of plasmids is only a metastable state. If $\Lambda_+ > \Lambda_0$, the arguments in \cite{assaf_fixation_2011,assaf_wkb_2017} suggest that the mean time until extinction grows roughly exponentially in the population size $N$. For bacteria in a bioreactor, where $N$ is often in the millions, we expect that extinction due to genetic drift will happen at timescales where other effects (such as \textit{de novo} mutations) become more prominent.

\section{Simulation details}

\label{apdx:simulation}

Since numerically simulating an exponentially growing population for long times is infeasible, in Figs.~\ref{fig:plasmid} and \ref{fig:plasmid_nonbinom} we simulated populations of fixed size $N$ throughout following a Moran process \cite{skanata_evolutionary_2016,jafarpour_evolutionary_2023} as described in the previous section, effectively a modification of the cloning algorithm \cite{giardina_direct_2006,lecomte_numerical_2007}. We estimated the population growth rate as
\begin{align}
    \hat \Lambda &= \frac{d(t)}{t} \log\left(\frac{N+1}{N}\right),
\end{align}

\noindent where $d(t)$ is the number of division events up to time $t$. To compute the backward distribution, we picked a random individual in the population at time $t$ and traced its ancestry. It can be shown that both the estimate for $\Lambda$ and the backward lineage converge to their true values in the limit $N, t \rightarrow \infty$ \cite{nemoto_finite-time_2017}. For points just before the phase transition, a small fraction of runs that became extinct due to stochastic fluctuations were discarded. To obtain the population distributions in Fig.~\ref{fig:plasmid_data}, we measured the number of plasmids in each cell present at time $t$.

In Fig.~\ref{fig:plasmid} we plot the fitness gradient with respect to $\log \beta$:
\begin{align}
    \sigma_{\log \beta} &= -\Lambda \E_b\left[ \frac{\partial \tau}{\partial \log \beta} \right] = -\beta \Lambda \E_b\left[ \frac{\partial \tau}{\partial \beta} \right] = \beta \sigma_\beta,
\end{align}

\noindent using \eqref{eq:selection_differential}, which differs from $\sigma_\beta$ by a smooth function of $\log \beta$.

\section{Derivation of asymptotic duality}

\label{apdx:duality_proof}

In this section we derive \eqref{eq:cgf_duality} by extending the ideas in \cite{pigolotti_generalized_2021,ocal_two-clock_2025}. Here we assume that the original quantities $n(t)$, $\log w(t)$ and $X(t)$, which all scale with $t$, satisfy a large deviation principle \cite{dembo_large_2010} of the form
\begin{align}
    p(n(t) = t \nu, \log w(t) = t \omega, X(t) = t \xi) &\sim e^{-t J_T(\nu, \omega, \xi)},
\end{align}

\noindent for a rate function $J_T$. Then using \cite{glynn_large_1994,duffy_how_2005} it can be shown that the variables $t_n$, $\log w_n$, and $X_n$ jointly satisfy a related large deviation principle of the form
\begin{align}
    p(t_n = n \tau, \log w_n = n \omega, X_n = n \xi) &\approx p\left(n(t) = \frac{t}{\tau}, w(t) =  t \, \frac{n(t)}{t} \, \omega, X(t) = t \, \frac{n(t)}{t} \, \xi\right) \sim e^{- n \tau J_T(1/\tau, \omega/\tau, \xi/\tau)},
\end{align}

\noindent with rate function
\begin{align}
    J_N(\tau, \omega, \xi) &:= \tau J_T(1/\tau, \omega/\tau, \xi/\tau),
\end{align}

\noindent which follows from the substitution $t := n \tau$. The rest of our derivation is a straightforward extension of the results in \cite{pigolotti_generalized_2021,ocal_two-clock_2025}. Varadhan's Lemma \cite{dembo_large_2010} allows us to estimate the behavior of the scaled cumulant generating function
\begin{align}
    \kappa_N(a,b,c) &= \lim_{n \rightarrow \infty} \frac 1 n \log \E_f \left[ e^{a t_n + b \log w_n + c X_n} \right] \label{eq:cgf_n}
\end{align}

\noindent in terms of the rate function as 
\begin{align}
    \kappa_N(a, b, c) &= \sup_{\tau,\omega,\xi} \tau a + \omega b + \xi c - J_N(\tau,\omega,\xi). \label{eq:legendre_n}
\end{align}

\noindent In other words, the cumulant generating function is the Legendre dual of the rate function. Similarly we can define
\begin{align}
    \kappa_T(a',b,c) &= \lim_{t \rightarrow \infty} \frac 1 t \log \E_f \left[ e^{a' n(t) + b \log w(t) + c X(t)} \right], \label{eq:cgf_t}
\end{align}

\noindent which is related to the rate function $J_T$ as
\begin{align}
    \kappa_T(a',b,c) &= \sup_{\nu,\omega,\xi} \nu a' + \omega b + \xi c - J_T(\nu,\omega,\xi). \label{eq:legendre_t}
\end{align}

\noindent Legendre duality then implies that $J_T$ is given by
\begin{align}
    J_T(\nu,\omega,\xi) &= \sup_{a',b,c} \nu a' + \omega b + \xi c - \kappa_T(a', b, c). \label{eq:legendre_t_dual}
\end{align}

With these prerequisites in place we can compute \eqref{eq:cgf_n} as 
\begin{align}
    \kappa_N(a,b,c) &= \sup_{\tau,\omega,\xi} \tau a + \omega b + \xi c - \tau J_T(1/\tau,\omega/\tau,\xi/\tau) = \sup_{\tau,\tilde \omega,\tilde \xi} \tau \left(a + \tilde \omega b + \tilde \xi c - J_T(1/\tau,\tilde \omega,\tilde \xi) \right),
\end{align}

\noindent where we substituted $\tilde \omega := \omega/\tau$, $\tilde \xi := \xi/\tau$. Now we use Legendre duality in \eqref{eq:legendre_t_dual} and obtain
\begin{align}
    \kappa_N(a,b,c) &= \sup_{\tau,\tilde \omega,\tilde \xi} \inf_{a',b',c'} \tau \left(a + \tilde \omega b + \tilde \xi c - \frac{a'}{\tau} - b' \tilde \omega - c' \tilde \xi + \kappa_T(a',b',c') \right).
\end{align}

\noindent We then use the minimax principle to swap the order of supremum and infimum and collect terms:
\begin{align}
    \kappa_N(a,b,c) &= \inf_{a',b',c'} \sup_{\tau,\tilde \omega,\tilde \xi} -a' + \tau \left(a + \tilde \omega (b - b') + \tilde \xi (c - c') + \kappa_T(a',b',c') \right).
\end{align}

\noindent If the term in brackets is nonzero, we can make the expression in the supremum arbitrarily large by multiplying it with a large $\tau$ of the same sign. At the infimum, therefore, this term must vanish. For the same reason, since we are free to choose $\tilde \omega$ and $\tilde \xi$, we must have $b = b'$ and $c = c'$ at the infimum, therefore 
\begin{align}
    a &= -\kappa_T(a', b, c).
\end{align}

\noindent This yields the duality relation
\begin{align}
    \kappa_N(a, b, c) = -a' &\quad\Leftrightarrow\quad \kappa_T(a', b, c) = -a, \label{eq:cgf_duality}
\end{align}

\noindent cf.~\cite{ocal_two-clock_2025}. We obtain \eqref{eq:psi_duality} from
\begin{align}
    \tilde \phi(\xi, \lambda) &= \kappa_T(-\xi, 1, \lambda), \\
    \tilde \psi(\alpha, \lambda) &= \kappa_N(-\alpha, 1, \lambda).
\end{align}

%Now since $\tau$ can be positive or negative, the supremum in the equation above can only be achieved if the term in brackets is $0$.
%TC:endignore
\end{document}